\documentclass[twocolumn]{aastex61}
\usepackage{amsmath,amstext}
\usepackage{apjfonts} 
\usepackage{afterpage}
\usepackage{color}
\usepackage[hang,flushmargin]{footmisc}

\usepackage{hyperref}
\usepackage{url}

\newcommand{\spitzer}{{\it Spitzer }}
\newcommand{\spitzerlong}{{\it Spitzer Space Telescope }}

\received{--}
\revised{--}
\accepted{--}

\shorttitle{Spitzer Observations of GW170817}
\shortauthors{Villar et al.}
\begin{document}

\title{Spitzer Space Telescope Infrared Observations of the Binary Neutron Star Merger GW170817}


\author{{V.~A.~Villar}\altaffilmark{1}, {P.~S.~Cowperthwaite}\altaffilmark{1}, {E.~Berger}\altaffilmark{1}, {P.~K.~Blanchard}\altaffilmark{1}, {S.~Gomez}\altaffilmark{1}, {K.~D.~Alexander}\altaffilmark{1}, {R.~Margutti}\altaffilmark{2}, {R.~Chornock}\altaffilmark{4}, {T.~Eftekhari}\altaffilmark{1}, 
{G.~G.~Fazio}\altaffilmark{1},
{J.~Guillochon}\altaffilmark{1}, {J.~L.~Hora}\altaffilmark{1}, {B.~D.~Metzger}\altaffilmark{3}, {M.~Nicholl}\altaffilmark{1}, \& {P.~K.~G.~Williams}\altaffilmark{1}}

\altaffiltext{1}{Harvard-Smithsonian Center for Astrophysics, 60 Garden Street, Cambridge, Massachusetts 02138, USA}
\altaffiltext{2}{Center for Interdisciplinary Exploration and Research in Astrophysics (CIERA) and Department of Physics and Astronomy, Northwestern University, Evanston, IL 60208}
\altaffiltext{3}{Department of Physics and Columbia Astrophysics Laboratory, Columbia University, New York, NY 10027, USA}
\altaffiltext{4}{Astrophysical Institute, Department of Physics and Astronomy, 251B Clippinger Lab, Ohio University, Athens, OH 45701, USA}

\begin{abstract}
We present \spitzerlong 3.6 and 4.5 \micron\ observations of the binary neutron star merger GW170817 at 43, 74, and 264 days post-merger. Using the final observation as a template, we uncover a source at the position of GW170817 at 4.5 $\micron$ with a brightness of $22.9\pm 0.3$ AB mag at 43 days and $23.8\pm 0.3$ AB mag at 74 days (the uncertainty is dominated by systematics from the image subtraction); no obvious source is detected at 3.6 $\micron$ to a $3\sigma$ limit of $>23.3$ AB mag in both epochs. The measured brightness is dimmer by a factor of about $2-3$ times compared to our previously published kilonova model, which is based on UV, optical, and near-IR data at $\lesssim 30$ days.  However, the observed fading rate and color ($m_{3.6}-m_{4.5}\gtrsim 0$ AB mag) are consistent with our model. We suggest that the discrepancy is likely due to a transition to the nebular phase, or a
reduced thermalization efficiency at such late time.  Using the \spitzer\ data as a guide, we briefly discuss the prospects of observing future binary neutron star mergers with \spitzer (in LIGO/Virgo Observing Run 3) and the {\it James Webb Space Telescope} (in LIGO/Virgo Observing Run 4 and beyond).
\end{abstract}


\section{Introduction} 
\label{sec:intro}

The gravitational wave discovery of the binary neutron star (BNS) merger GW170817 \citep{abbott2017gw170817}, and the subsequent identification of its electromagnetic counterpart \citep{abbott2017multi} provided the first multi-messenger view of a compact object merger and its aftermath. In the ultraviolet, optical, and near-infrared (hereafter, UVOIR) the emission was observed in the first month post-merger, before the source became sun-constrained \citep{andreoni2017,arcavi2017optical,SwopeDiscovery,cowperthwaite2017electromagnetic,diaz2017observations,drout2017light,evans2017swift,hu2017optical,kasliwal2017illuminating,lipunov2017master,pian2017spectroscopic,pozanenko2017grb170817a,smartt2017kilonova,DECamDiscovery,tanvir2017emergence,troja2017x,utsumi2017j,valenti2017discovery,villar2017combined}. This emission was produced by the radioactive decay of $r$-process nuclei synthesized in the merger ejecta, a so-called kilonova (e.g., \citealt{li1998transient,rosswog1999,metzger2010electromagnetic,roberts2011electromagnetic,metzger2012most,barnes2013effect,tanaka2013radiative}).

From these observations, most authors concluded that GW170817 produced at least two distinct non-relativistic ejecta components: a rapidly-evolving ``blue'' component dominated by light $r$-process nuclei (atomic mass number $A<140$) with a mass of $\approx 0.02$ M$_\odot$ and a velocity of $\approx 0.3c$; and a more slowly-evolving ``red'' component dominated by heavy $r$-process elements ($A>140$, including lanthanides) with a mass of $\approx 0.05$ M$_\odot$ and a velocity of $\approx 0.15c$ (e.g., \citealt{villar2017combined}; although see \citealt{smartt2017kilonova,waxman2017}). The multi-component nature of the ejecta is also evident in optical and NIR spectroscopic observations \citep{chornock2017electromagnetic,nicholl2017electromagnetic,pian2017spectroscopic,ShappeeSpectra,smartt2017kilonova}. Subsequently, X-ray, radio, and optical observations of the non-thermal afterglow provided insight into the production of relativistic ejecta \citep{alexander2017electromagnetic,gottlieb2017cocoon,haggard2017chandra,hallinan2017counterpart,lazzati2017jet,margutti2017electromagnetic,troja2017x,alexander2018,davanzo2018evolution,dobie2018turnover,lyman2018optical,margutti2018binary,mooley2018x,nynka2018fading,ruan2018x,troja2018x}.

At $\gtrsim 10$ days the kilonova spectral energy distribution (SED) peaked in the NIR, with a blackbody temperature of $\lesssim 1300$ K, and hence an expected substantial contribution into the mid-IR \citep{chornock2017electromagnetic,nicholl2017electromagnetic,kasliwal2017illuminating}.  Here, we present the full set of \spitzerlong IR observations of GW170817, obtained at 43, 74, and 264 days post-merger, which extend the kilonova observations to 3.6 and 4.5 $\micron$ (see \citealt{laugcn}); we uncover clear detections at 4.5 \micron.  In \autoref{sec:obs} we present the observations and our data analysis, image subtraction, and photometry procedures.  We compare the results to our kilonova models from \citet{villar2017combined} in \autoref{sec:dis}.  Motivated by the results, in \autoref{sec:bns_implications} we discuss the  prospects for IR observations of future events with \spitzer and the {\it James Webb Space Telescope} (JWST).

All magnitudes presented in this Letter are given in the AB system and corrected for Galactic reddening with $E(B-V) = 0.105$ mag \citep{schlafly11}. All uncertainties are reported at the $1\sigma$ level. We assume negligible reddening contribution from the host galaxy \citep{blanchard2017electromagnetic}, and a luminosity distance to NGC\,4993 of 40.7 Mpc \citep{cantiello2018precise}.

\section{Observations and Data Analysis}
\label{sec:obs}

\begin{figure*}[ht]
\centering
\includegraphics[width=\textwidth]{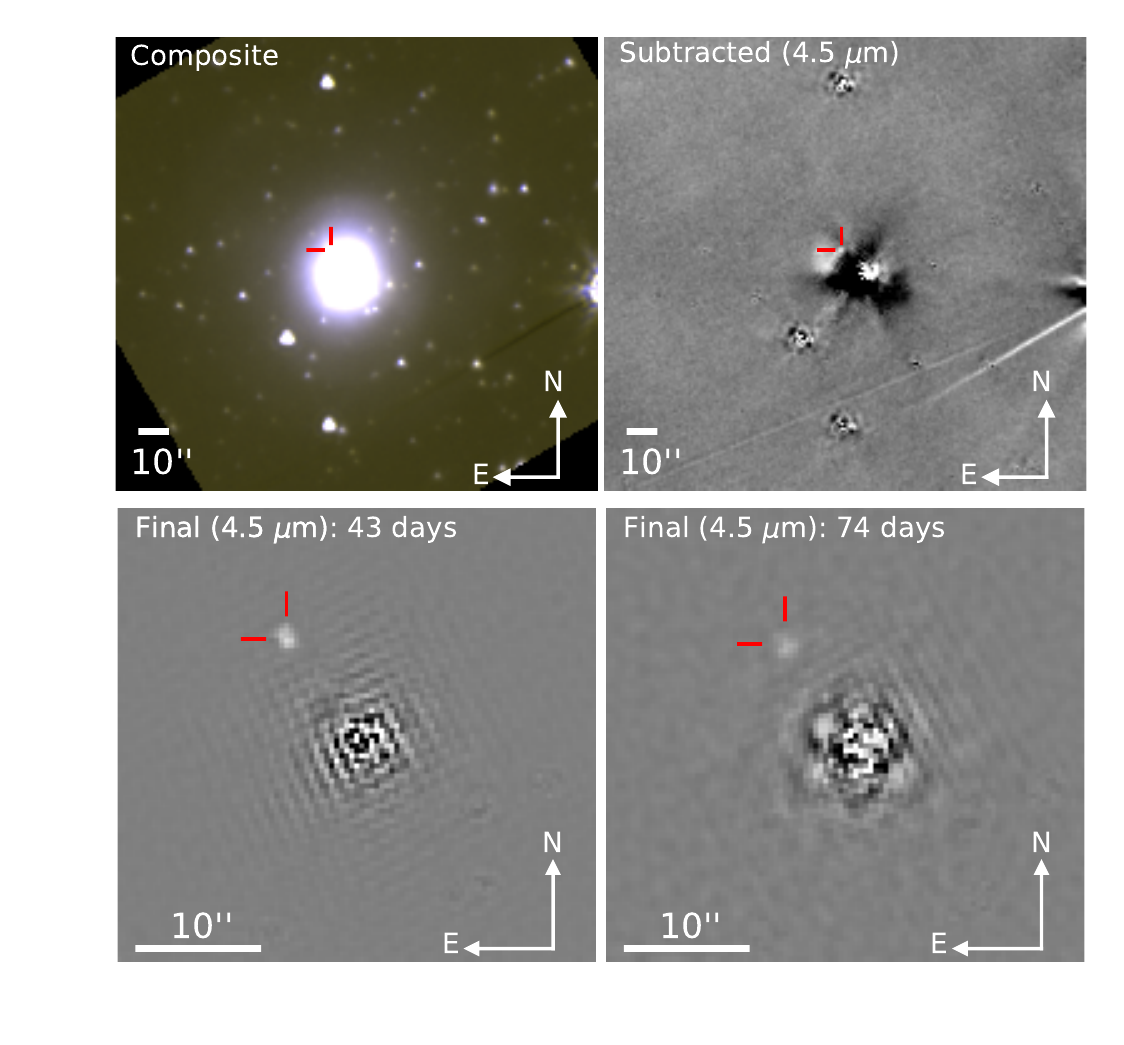}
\caption{{\it Upper Left:} \spitzer 3.6 and 4.5 $\micron$ color composite image from the observations on 2017 September 29 (43 days post-merger). The location of GW170817 is marked by the red cross hairs. {\it Upper Right:} Residual from {\tt HOTPANTS} image subtraction of the 4.5 $\micron$ images at 43 and 264 days. While a point source is visible at the position of GW170817, its location is contaminated by subtraction residuals from the host galaxy. {\it Bottom:} Residual images (Left: 43 days; Right: 74 days) after subtracting a masked and smoothed version of the image from itself (see Section~\ref{sec:obs}). An isolated point source is clearly visible at the location of GW170817.}
\label{fig:spitz}
\end{figure*}

We obtained public \spitzer \citep{werner2004spitzer} observations of GW170817 taken on 2017 September 29, 2017 October 30, and 2018 May 8 with the InfraRed Array Camera (IRAC; \citealt{fazio2004infrared}) in the 3.6 and 4.5 $\micron$ bands during the ``warm'' \spitzer mission (Director's Discretionary Time Program 13202; PI: Kasliwal); see Table~\ref{tab:photometry}. Each visit consisted of 466 frames with exposure times of 30 sec per frame, for a total on-source time of $\approx 3.9$ hours in each band.  We processed the images using standard procedures in {\tt Mopex} \citep{makovoz2005point} to generate mosaic images. {\tt Mopex} cleans the images of cosmic rays and applies appropriate distortion corrections before drizzling the images.  We used a drizzling factor of 0.8 and an output pixel scale of $0.4\arcsec$.  We compare the native astrometry to seven 2MASS point sources in the field, and find that the astrometric solution is good to about 1 pixel.

\begin{deluxetable}{cccc}
\label{tab:photometry}
\tablecolumns{4} 
\tablewidth{0pt} 
\tablecaption{ \spitzer/IRAC Observations and Photometry of GW170817} 
\tablehead{ 
\colhead{UT Date} & \colhead{Epoch} & \colhead{$\lambda$} & \colhead{Magnitude}\\
\colhead{} & \colhead{(days)} & \colhead{($\micron$)} & \colhead{}} 
\startdata 
2017 Sep 29 & 43 &  3.6 & $>23.3$\\
2017 Sep 29 & 43 &4.5 & $22.9\pm0.3$\\
2017 Oct 30 & 74 & 3.6 & $>23.3$\\
2017 Oct 30 & 74 &  4.5 & $23.8\pm0.3$\\
2018 May 8 & 264 &  3.6 & Template\\
2018 May 8 & 264 & 4.5 & Template
\enddata
\end{deluxetable}

We performed image subtraction with the {\tt HOTPANTS} package \citep{alard2000image,becker2015hotpants}, using the 2018 May 8 observations as a template in each band.  We note that at 264 days post-merger, the emission from the relativistic ejecta (which dominates in the radio and X-ray bands) has $m_{3.6}=25.9$ and $m_{4.5}=25.7$ mag \citep{alexander2018,margutti2018binary,xie2018}, more than an order of magnitude below the expected brightness of the kilonova emission, and well below the detection level of the \spitzer data.  A composite 3.6 and 4.5 $\micron$ image, and the subtracted 4.5 $\micron$ image at 43 days are shown in Figure~\ref{fig:spitz}.  A point source is apparent in the subtracted image.

Although {\tt HOTPANTS} computes and utilizes a spatially-variable convolution kernel, and is therefore able to match dissimilar point spread functions (PSFs), we find that the location of GW170817 is heavily contaminated by residual artifacts from the bright host galaxy. To remove the remaining contamination, we first mask the source location with a region the size of the expected PSF ($\approx 5$ pixels). We then smooth the masked image with a Gaussian kernel, interpolating across the masked region. We use a kernel standard deviation of one pixel (but find that the kernel width has little effect on our results). We then subtract the smoothed image from the original data to isolate the point source. The resulting final 4.5 $\micron$ images from 43 and 74 days are shown in Figure~\ref{fig:spitz} and clearly reveal the presence of a point source at the location of GW170817.

We measure the brightness of the source using both fixed aperture photometry and PSF-fitting assuming a Gaussian PSF. We injected fake point sources around the host galaxy at a similar offset to that of GW170817 to quantify the systematic uncertainties of the subtraction methods and photometry. For the observations with a detected source at the location of GW170817 (4.5 \micron), we specifically injected fake sources of the same measured magnitude. For each injected source, we executed the same method of smoothing and subtraction from a masked image. We used the spread in the recovered magnitudes as our overall uncertainty. For the observations without a significantly detected source (3.6 \micron), we injected sources with a range of fluxes to determine $3\sigma$ upper limits.  The results are summarized in Table~\ref{tab:photometry}. We additionally confirmed that the 4.5 \micron\ detection is not an artifact of the subtraction process or the IRAC PSF by searching for sources at the same relative location as the GW170817 counterpart around a nearby saturated star within the field of view, following the same procedure. We did not find any significant sources around the star.

\begin{figure*}[t]
\centering
\includegraphics[width=\textwidth]{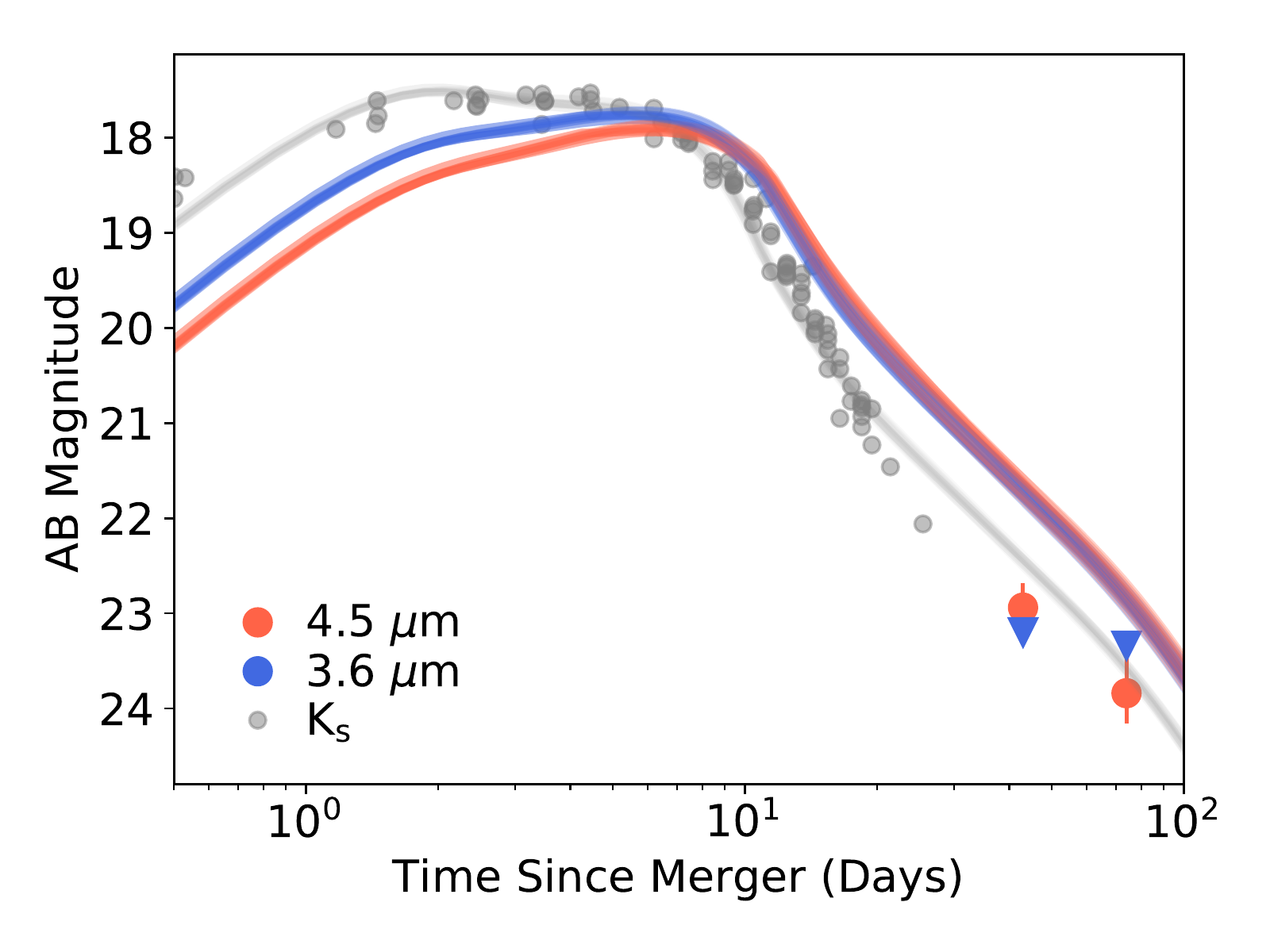}
\caption{\spitzer data at 3.6 \micron\ (blue) and 4.5 \micron\ (red); upper limits are marked as triangles. Also shown are light curves from our model fit to the complete UVOIR data set at $\lesssim 30$ days \citep{villar2017combined}.  For comparison we also show the $K_s$-band ($2.2$ \micron) data and model (gray; \citealt{cowperthwaite2017electromagnetic,drout2017light,kasliwal2017illuminating,smartt2017kilonova,tanvir2017emergence,troja2017x,utsumi2017j,villar2017combined}).}
\label{fig:lc}
\end{figure*}

The detected source at 4.5 $\micron$ has $22.9\pm 0.3$ mag at 43 days and $23.8\pm 0.3$ mag at 74 days post-merger. The source is detected with a signal-to-noise ratio (SNR) of $\approx 10$ at 43 days and $\approx 5$ at 74 days. However, the final uncertainties are dominated by systematic effects, as determined from the spread in magnitudes for the injected fake point sources. We do not detect a source at 3.6 $\micron$ in either epoch to a $3\sigma$ limit of $\gtrsim 23.3$ mag.

\section{Comparison to a Kilonova Model}
\label{sec:dis}

We compare the observations to our three-component kilonova model, which was previously used to fit all available UVOIR photometry \citep{villar2017combined}; see Figure~\ref{fig:lc}. Each component is characterized by a unique gray opacity roughly corresponding to its lanthanide fraction \citep{2018ApJ...852..109T}, and is independently described by a blackbody SED. The blackbody SEDs cool as a function of time until they reach a minimal ``temperature floor", at which point we assume that the photosphere recedes into the ejecta, at a constant temperature. At late times ($\gtrsim 10$ days), our three-component model predicts that the light curve is dominated by the intermediate $r$-process component and that this component has reached its temperature floor of $\approx 1300$ K, somewhat cooler than the lowest lanthanide ionization temperature (e.g., \citealt{kasen2013opacities}). 

We find that our model over-predicts the \spitzer measurements at 43 and 74 days by about a factor of $\approx 3$ (1.2 mag) and $\approx 2.5$ (1 mag), respectively (Figure~\ref{fig:lc}). However, the decline rate between the two measurements is in good agreement with the model prediction.  Similarly, the temperature implied by the flat or red color in the 3.6 and 4.5 $\micron$ bands ($\lesssim 1200$ K) is consistent with the temperature floor in our model.  We observe a similar late-time deviation from our model in the $K_s$-band (2.2 $\micron$) at $\gtrsim 20$ days. 

Assuming a blackbody SED with $T=1200$ K we find that the bolometric luminosity implied by the 4.5 $\micron$ detections is $\approx 6\times10^{38}$ erg s$^{-1}$ and $\approx 2\times10^{38}$ erg s$^{-1}$at 43 and 74 days, respectively. This is consistent with the drop off in bolometric luminosity starting at $\approx 10$ days, when the estimated bolometric luminosity is $\approx 2\times10^{40}$ erg s$^{-1}$ \citep{cowperthwaite2017electromagnetic,2018arcavi,waxman2017}

Relaxing some of the assumptions in our model may eliminate the brightness discrepancy. For example, at the time of the \spitzer observations the kilonova is likely transitioning into the nebular phase, and the blackbody SED approximation may break down. Using the parameters of the dominant intermediate-opacity component of our model \citep{villar2017combined}, we find that at 43 days the optical depth is $\tau\approx 1$, suggesting that the ejecta are becoming optically thin. Additionally, the shape of the late-time light curve is also dictated by the time-dependent thermalization efficiency of the merger ejecta \citep{barnes2016radioactivity}.  A steeper decline of the thermalization efficiency at $\gtrsim 20$ days will better capture the lower observed fluxes in the $K_s$ and \spitzer bands. The thermalization is highly dependent on the nuclear mass models assumed (see e.g., \citealt{rosswog2017detectability}), and is uncertain by  almost an order of magnitude at $\gtrsim 1$ month.

We also consider the possibility that the observed IR emission is due to reprocessing of bluer kilonova emission by newly formed dust. The warm temperature implied by our observations requires carbon-based dust, due to its high condensation temperature ($T_c\approx 1800$ K; \citealt{takami2014dust}). We fit a modified blackbody to the \spitzer photometry at day 43, assuming $m_{3.5}\approx m_{4.5}$ (following Equations 1 and 2 of \citealt{galldust}). We find that the carbon dust mass required to reproduce the observed luminosity is $\approx 5\times 10^{-7}$ M$_\odot$. However, \citet{galldust} explored a range of theoretical kilonova wind models and found that at most $\sim 10^{-9}$ M$_\odot$ of carbon dust can be produced. We therefore conclude that the observed IR emission is not due to dust reprocessing.

\section{Implications for IR Observations of Future BNS Mergers}
\label{sec:bns_implications}

The \spitzer detections of GW170817 at 43 and 74 days post-merger indicate that future BNS mergers should be observed at IR wavelengths.  Indeed, taking our models at face value, at least in the well-characterized regime at $\lesssim 20$ days, it appears that the peak of the kilonova emission shifts into the NIR/MIR bands at $\gtrsim 10$ days.  This suggests that significant effort should be focused on robust characterization of the IR emission of kilonovae. This will provide numerous benefits, including more accurate determination of the bolometric luminosity and therefore total $r$-process ejecta mass, improved measurements of the $r$-process opacity at long wavelengths, observational constraints on the late-time thermalization efficiency, and continued insight into BNS mergers as sites of cosmic $r$-process production. 

Advanced LIGO/Virgo (ALV) Observing Run 3 (O3) is expected to begin in early 2019 and span a full year, with an expected BNS merger detection distance of $\approx 120$ Mpc.  The timing of O3 overlaps favorably with \spitzer Cycle 14 (the final \spitzer cycle), and the sensitivity should be sufficient to detect events with a similar IR luminosity to GW170817 in the first $\approx 40$ days to $\approx 120$ Mpc.  For example, observations of about 9 hours on-source can achieve $5\sigma$ limiting magnitudes\footnotemark\footnotetext{See \href{http://ssc.spitzer.caltech.edu/warmmission/propkit/pet/senspet/}{\spitzer ETC}.} of $m_{3.6}\approx 25.5$ mag and $m_{4.6}\approx 25$ mag, assuming no significant contamination from host galaxy subtraction.

Beyond O3 (2021 and later), the ALV network is expected to achieve design sensitivity, with typical BNS merger detections to $\approx 200$ Mpc and a maximal detection distance of $\approx 450$ Mpc (for favorably oriented and positioned BNS mergers). The timing is ideal for overlap with JWST, which will be able to provide NIR and MIR spectra. In the NIR, NIRSpec can produce low-resolution ($R\approx 100$) spectra at $0.6-5.3$ $\micron$; this resolution is sufficient for kilonovae given the typical velocities of $\sim 0.1-0.3c$. In particular, spectra with ${\rm SNR}\gtrsim 50$ can be obtained near peak for a GW170817-like kilonova to $450$ Mpc in just 1 hour of on-source time.  At later times, BNS mergers could be tracked to $\approx 40$ days at $\approx 200$ Mpc with ${\rm SNR}\approx 10$ in about 6 hours of on-source time.

In the MIR, the Mid-Infrared Instrument (MIRI) can produce low-resolution ($R\approx 40$--$160$) spectra covering $5-14$ $\micron$. In particular, ${\rm SNR}\approx 10-20$ at $5-9$ \micron\ (and lower SNR at longer wavelengths) can be achieved for a GW170817-like kilonova near peak to $\approx 450$ Mpc with $\approx 5$ hours of on-source time. At late times ($\approx 40$ days), MIRI can produce ${\rm SNR}\approx 5$ spectra at $5-7$ \micron to $\approx 100$ Mpc. 

We do not yet know the full range of brightnesses and SEDs of kilonovae, as well as the potential contribution of dust reprocessing, but the discussion above illustrates that NIR/MIR characterization of kilonovae can be achieved with \spitzer in ALV O3 and with JWST when ALV reaches design sensitivity. This can be achieved with a modest time investment, but will require target-of-opportunity response to BNS mergers.

\section{Conclusions}
\label{sec:conc}

We present \spitzer IR observations of the kilonova associated with GW170817 spanning to 264 days post-merger.  We detect the kilonova at 4.5 $\micron$ at 43 and 74 days post-merger with a brightness of $\approx 22.9$ and $\approx 23.8$ mag, respectively.  We do not identify a confident detection at 3.6 $\micron$, to a $3\sigma$ upper limit of $\gtrsim 23.3$ mag. The inferred color of the kilonova indicates that the ejecta has cooled to $\lesssim 1200$ K at these late times. These magnitudes are fainter than an extrapolation of our model to the UVOIR data at $\lesssim 30$ days, highlighting the need for improved models at late times (for example, the details of the ejecta thermalization). Finally, we show that future BNS mergers with kilonovae similar to GW170817 will be detectable with \spitzer to 120 Mpc at 40 days post-merger, and will be accessible to NIR and MIR spectroscopy with JWST to $\approx 450$ Mpc at peak and to $\approx 100-200$ Mpc at 40 days post-merger (and to later times with JWST imaging).

\acknowledgements
The Berger Time-Domain Group at Harvard is supported
in part by the NSF through grant AST-1714498, and by
NASA through grants NNX15AE50G and NNX16AC22G. This work is based in part on observations made with the {\it Spitzer Space Telescope}, which is operated by the Jet Propulsion Laboratory, California Institute of Technology under a contract with NASA.

\bibliography{mybib.bib}

\end{document}